\documentclass[journal]{IEEEtran}

\usepackage{amsmath}
\usepackage{amssymb}
\usepackage{amsthm}

\newtheorem{definition}{Definition}

\newtheorem{remark}{Remark}

\usepackage{graphicx}
\usepackage{csquotes}
\usepackage{multirow}
\usepackage{booktabs}
\usepackage{amsmath}
\usepackage[colorinlistoftodos]{todonotes}
\usepackage{amsfonts}
\usepackage{tikz,pgfplots}
\usepackage{caption}
\usepackage{subcaption}
\usepackage{pgfplots}
\usepackage{optidef}

\usepgfplotslibrary{fillbetween}
\pgfmathdeclarefunction{gauss}{2}{%
  \pgfmathparse{1/(#2*sqrt(2*pi))*exp(-((x-#1)^2)/(2*#2^2))}%
}

\usepackage{booktabs}
\usepackage{array}
\newcolumntype{L}[1]{>{\raggedright\arraybackslash}p{#1}}
\usepackage{multirow}
\usepackage{todonotes}

\DeclarePairedDelimiter\floor{\lfloor}{\rfloor}

\usepackage{csvsimple}
\usepackage{dsfont}

\usepackage{blindtext}
\usepackage[linesnumbered,boxed]{algorithm2e}

\usepackage[hyphens]{url}

\SetAlCapSkip{1em}
\title{Achievable Rates of Nanopore-based DNA Storage}

\author{Brendon McBain and Emanuele Viterbo\\
ECSE Dept., Monash University, Clayton, Australia \\
Email: \{brendon.mcbain,\;emanuele.viterbo\}@monash.edu}

\begin{document}

\maketitle

\begin{abstract}
    This paper studies achievable rates of nanopore-based DNA storage when nanopore signals are decoded using a tractable channel model that does not rely on a basecalling algorithm. {Specifically, the noisy nanopore channel (NNC) with the Scrappie pore model generates average output levels via i.i.d. geometric sample duplications corrupted by i.i.d. Gaussian noise (NNC-Scrappie).} Simplified message passing algorithms are derived for efficient soft decoding of nanopore signals using NNC-Scrappie. Previously, evaluation of this channel model was limited by the lack of DNA storage datasets with nanopore signals included. This is solved by deriving an achievable rate based on the dynamic time-warping (DTW) algorithm that can be applied to genomic sequencing datasets subject to constraints that make the resulting rate applicable to DNA storage. Using a publicly-available dataset from Oxford Nanopore Technologies (ONT), it is demonstrated that coding over multiple DNA strands of $100$ bases in length and decoding with the NNC-Scrappie decoder can achieve rates of at least $0.64-1.18$ bits per base, depending on the channel quality of the nanopore that is chosen in the sequencing device per channel-use, and $0.96$ bits per base on average assuming uniformly chosen nanopores. These rates are pessimistic since they only apply to single reads and do not include calibration of the pore model to specific nanopores.
\end{abstract}

\section{Introduction}

DNA storage is a promising technology for densely archiving large amounts of data in DNA molecules. However, reading back the archived data requires a biosensor to detect the DNA molecules that is often prone to errors. These limit the storage efficiency since error-correction codes with redundancy are required for reliable DNA storage \cite{Milenkovic2024,Sabary2024}. A popular choice for this biosensor is the nanopore, in particular the Oxford nanopore used in the DNA sequencers by {\em Oxford Nanopore Technologies (ONT)}. While these nanopore sequencers offer many practical benefits, such as affordability and portability, they suffer from severe synchronisation errors that greatly impede our ability to achieve efficient and reliable nanopore-based DNA storage. With this in mind, tractable models of the nanopore are required to characterise optimal codes using information-theoretic techniques and provide theoretical guarantees on the achievable storage efficiency \cite{BITSDNA}.

The nanopore is a microscopic pore that is used as a biosensor to detect nucleotide molecules (or ``bases'') $\{\mathsf{A},\mathsf{T},\mathsf{C},\mathsf{G}\}$ that compose single-stranded DNA (ssDNA). The nanopore is equipped with a membrane that allows the flow of electrical current across the pore with a response depending on the amount of blockage caused by the bases passing through it. The base-sequence is passed through the nanopore using a motor protein that attempts to shift in and hold a sequence of bases inside it for a sufficiently long enough time to sample the current level. However, the motor protein shifts in bases at non-uniform time intervals, while the sampling of the current response occurs at a constant sample rate of $f_s = 4$kHz. This results in a variable number of samples per base (sample duplications), making reliable detection of the base-sequence a challenging synchronisation problem. In addition, the current samples are corrupted by measurement noise that includes chemical noise, due to random variations in the concentration of molecules in the pore, and thermal noise. In nanopore sequencers, the nanopores reside in a {\em flowcell} that contains a number of channels that hold $4$ different nanopores. That is, if there are $512$ channels then there are $2{,}048$ nanopore channels that may be chosen to read a DNA strand. When considering data from a nanopore sequencer, we must differentiate between the raw nanopore channels and the flowcell channel that chooses the nanopore to use per channel use. 

There are three prominent models for nanopore sequencing in the literature \cite{BITSDNA}:

\subsubsection{{Pore Model with Backtracking and Skipping}}

In \cite{Laszlo2014}, a {\em pore model} was found for a non-Oxford nanopore by estimating the average levels in the raw current signal depending on the $\tau$ bases inside the nanopore. In \cite{Mao2018}, this pore model was used to propose a channel model of the nanopore sequencer for an information-theoretic analysis. The sampling rate was $f_s = 50$kHz ($\times 10$ higher than in current-day Oxford nanopore sequencers), allowing the sample duplications to essentially be eliminated using a segmentation algorithm and other distortions in the channel to come to the forefront. In particular, the model was centred around backtracking (the same sub-sequence of bases pass through the pore again) and skippings/deletions (missing a sample for sub-sequences of bases that passed too quickly through the pore). However, backtrackings were later ignored---since they were relatively infrequent in \cite{Laszlo2014}---and instead focused on a deletion channel with memory, which had only been analysed in the memoryless case. In addition, this channel included a random fading interval around the average levels in the pore model and discretised the samples. 

\subsubsection{{Basecaller Model with Insertions, Deletions, and Substitutions}}
The typical approach to modelling the nanopore sequencer, and DNA sequencers in general, is to include a basecaller neural-network that estimates the base-sequence from the raw current signal. By counting the rate of insertions, deletions, and substitutions in the basecalled output, the classical {\em insertion-deletion-substitution (IDS)} channel can be parameterised. In \cite{Hamoum2021,Hamoum2023}, the IDS channel was extended to include memory between IDS errors to capture the effect of the memory in the nanopore channel on the basecalled outputs. The IDS channel {\em with memory} greatly improves upon its memoryless counterpart for 
nanopore sequencing.

\subsubsection{{Pore Model with Noisy Duplications}}
In \cite{McBain2022,McBain2024a}, the pore model approach taken in \cite{Mao2018} was analogously extended to Oxford nanopores. Since the sampling rate of Oxford nanopore sequencers is $f_s=4$kHz (recently increased to $5$kHz), the sample duplications play a much more important role than the backtracking and skippings in \cite{Mao2018}, since fewer samples per level greatly increases the difficulty of accurate segmentation \cite{BITSDNA}. Therefore, channels with memoryless noisy duplications and memory between the inputs were proposed for modelling the nanopore channel. The memory between inputs allows the inclusion of a pore model that models the nanopore memory, and the noisy duplications model the duplications of samples in the presence of measurement noise. When the state-space is a de-Bruijn graph, the channel is commonly referred to as the {\em noisy nanopore channel (NNC)}. This channel is the focus of this paper, with geometrically-distributed sample duplications and AWGN measurement noise.

In general, pore{-based} models are more challenging to parameterise than basecaller{-based} models since they require learning at least $4^\tau$ parameters ($\tau \sim 5$ for the R9.4.1 Oxford nanopore and $\tau \sim 10$ for the latest R10.4 Oxford nanopore). However, pore{-based} models are potentially more accurate since they { directly} decode the information-rich raw nanopore signals. { It remains unclear which approach offers the greatest potential for efficient and reliable nanopore-based DNA storage. Therefore, pursuing both strategies in parallel is valuable as the technology matures. To this end}, this paper will demonstrate that low-complexity decoders based on existing pore models can be very efficient in terms of information rate. Now, let us give a background on the coding and information-theoretic results for all of these synchronisation channels.

In \cite{Mao2018}, Dobrushin's classical coding theorem for the i.i.d. deletion channel was extended to include memory over the channel inputs so that the pore model could be used with skipping errors of levels that have memory. Recently, codes for correcting the backtracking and skipping errors were proposed \cite{Chee2024}. In \cite{Morozov2024}, Dobrushin's coding theorem for the i.i.d. IDS channel was extended to include Markov insertions, deletions and substitution errors as in the IDS channel for nanopore sequencing \cite{Hamoum2023}. In \cite{Maarouf2023}, numerical achievable rates were computed for the IDS channel of nanopore-based DNA storage for coding over multiple short base-sequence inputs. Remarkably, the capacity of these IDS channels can be realised with efficient encoding/decoding using the polar codes in \cite{Pfister2021}. In \cite{McBain2024b}, the coding theorem was proven in the setting of Markov-constrained inputs of noisy duplication channels with i.i.d. noisy duplications in levels with memory as in the pore model. In \cite{McBain2022}, numerical achievable rates were computed for the NNC parameterised by the Scrappie nanopore simulator (NNC-Scrappie) \cite{Scrappie}. In \cite{Rameshwar2024}, closed-form capacity bounds for general NNCs were proposed. A concatenated coding scheme was developed for the NNC in the finite block-length regime based on minimum-distance decoding with the {\em dynamic time-warping (DTW)} distance \cite{Vidal2023,Vidal2023_b,Vidal2024}. A homophonic concatenated coding scheme for NNCs was proposed in \cite{McBain2023,McBain2025a}, which included an inner DNA source constraint code such as the one proposed in \cite{Benerjee2021}.


In this paper, an information-theoretic method is developed to compute numerical achievable rates from raw nanopore signals from nanopore channels when decoding using the NNC-Scrappie channel model. In particular, we consider the DNA storage scenario where coding is done over multiple short base-sequences. This is particularly challenging for channel models based on pore models since the only suitable datasets that are available in the public domain include variable-length genomic base-sequences rather than short, fixed-length synthetic base-sequences as in DNA storage. A significant contribution of the method in this paper is its ability to compute achievable rates for DNA storage from genomic datasets. In addition, the message passing algorithms used in the NNC-Scrappie decoder from \cite{McBain2024a} are simplified and made more accurate in the case of i.i.d. geometric duplications. This simplification makes their implementation particularly easy, and are readily used by bioinformaticists for applications beyond DNA storage. Using this decoder, we compute achievable rates for $10$ different raw nanopore channels in a flowcell from the ONT dataset \cite{ONTdataset} for the R9.4.1 Oxford nanopore. These numerical results characterise the variation in channel quality depending on the choice of nanopore channel per channel-use. Assuming a randomly chosen nanopore channel, the achievable rate is approximately $0.96$ bits/base with a degradation of $0.18$ bits/base compared to the ideal NNC-Scrappie channel. This provides numerical evidence on the accuracy of the NNC-Scrappie channel model proposed in \cite{McBain2024a}. Moreover, this accuracy is likely pessimistic and could be improved upon with calibration to individual nanopores in the flowcell.

\section{Preliminaries}

\subsection{Notation}
A vector of length $m$ is written as $x_1^m = (x_1,x_2,\ldots,x_m)$. For brevity, the vector $x_1^m$ may instead be written as $\mathbf{x}$ whose length is given by $\mathsf{len}(\mathbf{x})$. The concatenation of vectors $\mathbf{x}$ and $\mathbf{y}$ is $\mathbf{x} \circ \mathbf{y} = (x_1,\ldots,x_{\mathsf{len}(\mathbf{x})}, y_1,\ldots, y_{\mathsf{len}(\mathbf{y})})$. The number of occurrences of $b$ in the vector $\mathbf{b}$ is denoted by $N_b(\mathbf{b})$. The probability of a realisation $x$ for random variable $X$ is $\mathbb{P}(X=x)$, written short-hand as $p(x) = \mathbb{P}(X=x)$ unless another probability measure is specified. The expectation of a joint random variable $(X,Y)$ is written as $\mathbb{E}[f(X,Y)]$ for any measurable function $f$. If a probability measure $(U\times W)(x,y) = U(x)W(y|x)$ is used instead of $\mathbb{P}(X=x,Y=y)$, then the expectation is written as $\mathbb{E}_{U \times W}[f(X,Y)]$.

\subsection{Nanopore Channel Model}

\subsubsection{Noiseless Nanopore Channel}
The input to the nanopore sequencer is the base-sequence $b_1^n \in \mathcal{B}^n$ for bases $\mathcal{B}=\{\mathsf{A},\mathsf{T},\mathsf{C},\mathsf{G}\}$, where a sub-sequence $b_1^\tau \in \mathcal{B}^\tau$ of the input is called a {\em $\tau$-mer}. The output level of the nanopore sequencer without noise is determined by the function $f: \mathcal{B}^\tau \rightarrow \mathbb{R}$ that is specified by mapping $f(b_1^\tau)=x$ for input $b_1^\tau$ and level $x$. Moreover, the input sequence can represented as a sequence of $\tau$-mer sequences $s_1^m$, which we call the channel states, by taking a sliding window of length $\tau$ such that $s_{\ell}=b_{\ell}^{\ell + \tau - 1}$ for all $\ell$, $1\leq \ell \leq m = n - \tau + 1$. Since each $s_\ell$ only depends on $s_{\ell-1}$, the input to the sequencer can be represented as a first-order Markov chain on state-space $\Omega$ of $4^\tau$ states with transition matrix $P=[p(s|s')]$. Conditioned on an initial channel state $q\in\Omega$, the channel states belong to $\mathcal{S}_q = \{s_1^m: p(s_1|q)p(s_2|s_1)\cdots p(s_m|s_{m-1}) > 0\}$. The channel states correspond to the output levels $f(s_1^m) = (f(s_1), f(s_2), \ldots,f(s_m)) = x_1^m$.

\subsubsection{Noisy Nanopore Channel}
While the input base-sequence is driven through the nanopore by a motor, the output levels are sampled at a constant sampling rate $f_s$. Since the motor cannot maintain a constant speed, the output level $x_\ell$ for $\tau$-mer $s_\ell$ is duplicated such that there are $k_{\ell}$ identical sample levels $f(s_{\ell},s_{\ell},\ldots,s_{\ell})=(x_{\ell},x_{\ell},\ldots,x_{\ell})=\mathsf{stretch}(s_{\ell},k_{\ell})$. Then, in terms of samples, the channel states are the sample channel states
\begin{align}
z_1^{t_m}&= \mathsf{stretch}(s_1^m, t_1^{m})\notag\\
&=(\underbrace{s_1,\ldots,s_{1}}_{k_1}, \underbrace{s_{2},\ldots,s_{2}}_{k_2}, \ldots, \underbrace{s_{m},\ldots,s_{m}}_{k_m})
\end{align}
where $t_{\ell} = \sum_{i\leq \ell}k_i$ is the $\ell$-th jump time that corresponds to the sample where a change from channel state $s_{\ell}$ to $s_{\ell+1}$ occurs, resulting in a jump in the output levels. The sequence of jump times, which we refer to as the {\em segmentation}, belongs to the set $\mathcal{T}_{m} = \cup_{t_m\in \mathbb{N}} \mathcal{T}_{m,t_m}$ where $\mathcal{T}_{m,t_m} = \{t_1^{m}: t_m=\sum_{i=1}^{m}k_i\text{ and }t_\ell -t_{\ell-1} > 0\}$ is the set of all segmentations conditioned on $t_m$ total samples. Finally, each sample level $f(z_t)$ is corrupted by measurement noise $n_t$ to give the noisy sample levels $y_t = f(z_t) + n_t$ for all $t$.

The noisy nanopore channel model is now defined for modelling the nanopore sequencer. This channel models the durations $k_1^{m}$ as i.i.d. geometric random variables on $\mathbb{N}_{>0}$ (the positive integers) with a finite mean $\mathbb{E}[K]$. The noise values $n_1^{t_m}$ are modelled as i.i.d. zero-mean Gaussian random variables with variance $\sigma^2$. For notational simplicity, the channel input is in terms of the channel states rather than the base-sequence. The set of the number of possible output observations is $\mathbb{N}_{\geq m} =\{t\in \mathbb{N}: t \geq m\}$.

\begin{definition}[Noisy nanopore channel]\label{def:nnc}
    The noisy nanopore channel $W(\cdot|\cdot,q) : \mathcal{S}_{q} \rightarrow \cup_{t\in \mathbb{N}_{\geq m}}\mathbb{R}^{t}$, for initial state $q\in \Omega$, is defined by the channel transition probability
\begin{align}\label{eq:W}
    W(\mathbf{y}|\mathbf{s},q) &= C' \sum_{\mathbf{t}\in\mathcal{T}_{m,t_m}} e^{-\frac{1}{2\sigma^2}||\mathbf{y} - \mathsf{stretch}(f(\mathbf{s}),\mathbf{t})||^2} 
\end{align}
with $C' = (2\pi \sigma^2)^{-\frac{t_m}{2}} (1-\frac{1}{\mathbb{E}[K]})^{t_m - m} (\frac{1}{\mathbb{E}[K]})^m$, for all inputs $\mathbf{s} \in \mathcal{S}_{q}$ and all channel outputs $\mathbf{y}\in \cup_{t\in \mathbb{N}_{\geq m}} \mathbb{R}^t$. 
\end{definition}

Throughout this paper, we assume that the NNC is driven by an identically and uniformly distributed (i.u.d.) source as $U(\cdot|q) \sim \mathsf{Unif}(\mathcal{S}_q)$. This corresponds to a Markov source that chooses base inputs uniform randomly. Hence, the joint input-output channel process obeys $(U \times W)(\mathbf{s},\mathbf{y}|q)=U(\mathbf{s}|q) W(\mathbf{y}|\mathbf{s},q)=W(\mathbf{y}|\mathbf{s},q)/4^m$.

\subsection{DTW}

A fundamental distance measure for the noisy nanopore channel is the DTW distance~\cite{Webb2021TightLB}. For state level sequence $\mathbf{x}=f(\mathbf{s})$ and channel output $\mathbf{y}$, DTW distance minimises the Euclidean distance between $\mathsf{stretch}(\mathbf{x},\mathbf{t})$ and $\mathbf{y}$ with respect to the segmentation $\mathbf{t}$. In general, DTW allows for insertions (which, when applied to $\mathbf{x}$, are duplications) and deletions, however here we consider a special case that only allows duplications. In particular, we have DTW distance
\begin{align}
    d_{\mathsf{dtw}}(\mathbf{x},\mathbf{y}) = \min_{\mathbf{t}\in\mathcal{T}_{m,t_m}} || \mathbf{y} - \mathsf{stretch}(\mathbf{x},\mathbf{t}) ||^2 
\end{align}
and normalised DTW distance 
\begin{align}
    \sigma_{\mathsf{dtw}} = \sqrt{\frac{d_{\mathsf{dtw}}(\mathbf{x},\mathbf{y})}{\mathsf{len}(\mathbf{y})}} .
\end{align}
Since we consider the case of AWGN and geometric duplications, resulting in a log-additive likelihood of the joint input-output channel process with respect to time, the DTW distance is a maximum-likelihood estimate of the segmentation \cite{Jackson2003}. This is observed in (\ref{eq:W}) if we consider a maximisation of the negative log-likelihood of each segmentation, which is to maximise each term in the summation, giving a minimisation of Euclidean distance for each segmentation (with some constants that do not change the segmentation that achieves this minimum). Moreover, the normalised DTW distance is an asymptotic lower bound on the measurement noise $\sigma$ given by
\begin{align}
    \sigma \geq \liminf_{m\rightarrow\infty}\sigma_{\mathsf{dtw}}(\mathbf{X},\mathbf{Y})
\end{align}
almost surely with $\mathbf{Y}|\mathbf{S},q \sim W$ for initial state $q$ and $\mathbf{X}=f(\mathbf{S})$, where $W$ is the noisy nanopore channel with AWGN noise level $\sigma$. This bound arises since DTW distance minimises the mean-square error over all possible segmentations, including the true one. Indeed, the true segmentation would give an asymptotic mean-square error of $\sigma^2$.

DTW distance can be efficiently computed using the DTW algorithm, which, in our case, is based on the dynamic program 
\begin{IEEEeqnarray}{rCl}
D_{m,t_m}
&=& \bigl(y_{t_m} - f(s_{m})\bigr)^2
\nonumber\\
&& {} + \min\bigl\{
D_{m, t_m - 1},
\, D_{m-1, t_m - 1}
\bigr\}.
\label{eq:DTW_algorithm}
\end{IEEEeqnarray}

This dynamic program will be the backbone of the message passing algorithms to come. In addition, many useful approximations are available to reduce the complexity of the DTW algorithm, making it very efficient in practice~\cite{Webb2021TightLB}. 

\section{Message Passing Algorithms}

In this section, efficient message passing algorithms are developed for computing {\em a posteriori probabilities (APPs)} for the noisy nanopore channel in Definition \ref{def:nnc}. These algorithms will be a special case of the algorithms in \cite{McBain2022}, which were derived for arbitrary measurement noise and duplication processes for general noisy duplication channels. As a result, the message passing algorithms derived here are greatly simplified.

For all initial states $q \in \Omega$, channel inputs $\mathbf{s}\in \mathcal{S}_{q}$, and channel outputs $\mathbf{y} \in \cup_{t\in \mathbb{N}_{\geq m}} \mathbb{R}^t$, the APP is defined as 
\begin{align}
    V(\mathbf{s}|\mathbf{y},q) &= \frac{W(\mathbf{y}|\mathbf{s},q)}{\sum_{\mathbf{s}'\in\mathcal{S}_{q}}W(\mathbf{y}|\mathbf{s}',q)}.
\end{align}
The challenge in efficiently computing the APP $V(\mathbf{s}|\mathbf{y},q)$ is that it requires computing $W(\mathbf{y}|\mathbf{s},q)$ as in (\ref{eq:W}) which includes a summation over all segmentations in $\mathcal{T}_{m,t_m}$ for $m$ inputs and $t_m$ output samples. Clearly, a brute-force algorithm is not feasible even for moderate values of $m$ since it has combinatorial complexity. Instead, we derive dynamic programs to compute the numerator and denominator of the APP. Then, we will show how to implement these dynamic programs while ensuring numerical stability and efficient memory usage. 

\subsection{Trellis Representation}
The key idea for deriving efficient message passing algorithms will be the trellis representation of the sample state process with respect to time $t$. Due to the duplications, each sample state can be fully described using a state $s\in\Omega$, which specifies the bases inside the nanopore, and the segment index $\ell \in \{1,2,\ldots,m\}$, which specifies the ordering of this level state. Therefore, the sample states form a Markov chain with transition probabilities

\begin{align}
    \begin{split}
        p(\ell,s| \ell', s') &= \begin{cases} 
       \frac{p(s|s')}{\mathbb{E}[K]} &  \ell=\ell'+1, s\neq s'\\
        1-\frac{1}{\mathbb{E}[K]}& \ell=\ell', s = s'\\
      0 & \text{otherwise}
   \end{cases}
   \end{split}
\end{align}
for all $\ell,\ell' \in \{1,2,\ldots,m\}$ and for all $s,s' \in \Omega$. The first case is a ``jump'' which requires that the segment index increases from $\ell'$ to $
\ell=\ell'+1$ and the level state changes from $s'$ to $s$ with $s' \neq s$. The second case is a ``sample duplication'' which requires that the segment index and level state remain unchanged as $\ell=\ell'$ and $s=s'$. Since the level states change with respect to the segment index, there is no contribution from the state transition probability when a duplication occurs. In addition, the trellis representation based on $p(\ell,s| \ell', s')$ must additionally enforce a {\em segmentation constraint} that ensures when $t = t_m = \mathsf{len}(\mathbf{y})$ we have $\ell=m$, since the number of observation samples and the number of segments is known at the decoder.

We note that since the duplication probability is a constant $1-\frac{1}{\mathbb{E}[K]}$ across time, we do not require extra states to specify how many duplications have previously occurred as was required in \cite{McBain2022} for general duration distributions. For the general case, the sample states are a semi-Markov state process that can be analogously embedded in a Markov chain \cite{McBain2024b}.

\subsection{Derivation of Dynamic Programs}

Let $\alpha_{t}(\ell,s) = p(y_1^t,s,\ell|q)$ be the probability of all samples up to $t$ being $y_1^t$ conditioned on the segment index $\ell$ and segment state $s$. Let $\alpha_{t}(\ell) = p(y_1^t,s_1^{\ell}|\ell,q)$ be the probability of all samples up to $t$ being $y_1^t$ with segment states $s_1^{\ell}$ conditioned on the segment index $\ell$. Note that we ignore the constants in the following derivations, since they result in $C'/4^m$ in both the numerator and the denominator of the APP. The forward probabilities of the observations are
\begin{IEEEeqnarray}{rCl}
\alpha_{t}(\ell,s) 
&=& p(y_t,s,\ell \mid y_1^{t-1},s,\ell) \, p(y_1^{t-1},s,\ell)
\nonumber\\
&& {} + p(y_t,s,\ell \mid y_1^{t-1},\ell-1)\, p(y_1^{t-1},s',\ell-1)
\notag\\
&=& p(y_t \mid s,\ell)\Bigl[
p(s,\ell \mid s,\ell)\, \alpha_{t-1}(\ell,s)
\nonumber\\
&& {} + \sum_{s' \in \Omega} p(s,\ell \mid s',\ell-1)\, \alpha_{t-1}(\ell-1,s')
\Bigr]
\notag\\
&\propto& \exp\biggl(-\frac{1}{2\sigma^2}\bigl(y_t - f(s)\bigr)^2\biggr)
\nonumber\\
&& {}\times \Bigl[
\alpha_{t-1}(\ell,s)
+ \sum_{s' \in \Omega} \alpha_{t-1}(\ell-1,s')
\Bigr]
\label{eq:FA}
\end{IEEEeqnarray}
where the base case is $\alpha_{0}(0,q) = 1$ and zero otherwise. The forward probabilities of the observations and segment states are
\begin{IEEEeqnarray}{rCl}
\alpha_{t}(\ell)
&=& p(y_t, s_{\ell}, \ell \mid y_1^{t-1}, s_{\ell}, \ell)\, p(y_1^{t-1}, s_{\ell}, \ell)
\nonumber\\
&& {} + p(y_t, s_{\ell}, \ell \mid y_1^{t-1}, \ell-1)\, p(y_1^{t-1}, s_{\ell-1}, \ell-1)
\notag\\
&=& p(y_t \mid s_{\ell}, \ell)\Bigl[
p(s_{\ell}, \ell \mid s_{\ell}, \ell)\, \alpha_{t-1}(\ell)
\nonumber\\
&& {} + p(s, \ell \mid s_{\ell-1}, \ell-1)\, \alpha_{t-1}(\ell-1)
\Bigr]
\notag\\
&\propto& \exp\biggl(-\frac{1}{2\sigma^2}\bigl(y_t - f(s_{\ell})\bigr)^2\biggr)
\nonumber\\
&& {}\times \Bigl[
\alpha_{t-1}(\ell)
+ \alpha_{t-1}(\ell-1)
\Bigr]
\label{eq:CFA}
\end{IEEEeqnarray}
where the base case is $\alpha_{0}(0,q) = 1$ and zero otherwise. Therefore, the APP of $s_1^m$ is
\begin{align}
    V(\mathbf{s}|\mathbf{y}, q) 
    &= \frac{\alpha_{t_m}(m)}{\sum_{s \in \Omega} \alpha_{t_m}(m,s)} 
\end{align}
which can be efficiently computed using the derived dynamic programs in (\ref{eq:FA}) and (\ref{eq:CFA}).

\subsection{Implementation}

Now let us consider the implementation of the dynamic programs in (\ref{eq:FA}) and (\ref{eq:CFA}) for the {\em forward algorithm} and {\em conditional forward algorithm}, respectively. Firstly, we consider the log-probabilities $\log \alpha_t(\ell,s)$ and $\log \alpha_t(\ell)$. This has the purpose of ensuring numerical stability by using log semi-ring operations in the log domain (multiplication becomes addition, and addition becomes the LogSumExp ($\mathsf{LSE}$) function of two elements). In addition, it transforms the dynamic programs to a form almost identical to the dynamic program of the DTW algorithm, where the only significant difference is that the marginalisation of the ``jump'' and ``duplication'' events in (\ref{eq:CFA}), through the sum inside the square brackets, becomes $\min$ in (\ref{eq:DTW_algorithm}). This allows us to implement the dynamic programs as the DTW-inspired pseudo-code in Algorithm \ref{alg:FA} and Algorithm \ref{alg:CFA}. A key difference between these algorithms is that we do not need to store the log-probabilities at $(\ell,s)$ for all $\ell$, since at the $\ell$-th iteration we only require the result of the $(\ell-1)$-th iteration and once $\ell=m$ the algorithm terminates and returns the final result. Both algorithms have a time-complexity $O(mt_m|\Omega|)$, which is $O(m^2\mathbb{E}[K]|\Omega|)$ on average; observe that the constants of these time-complexities include the average duration and the number of states, thus these parameters have a considerable influence on the run-time in practice.

\setlength{\algomargin}{2em}

  \begin{algorithm}
    \SetAlgoLined
    \SetKw{Is}{is}
    \SetKw{Not}{not}
    \SetKw{In}{in}
    \SetKw{Continue}{continue}

    \tcp{Initialisation}
    $F, F' \leftarrow - {\infty}_{|\Omega| \times \mathsf{len}(\mathbf{y})}$\\
    \For{$n \in \{1,2,\ldots,\mathsf{len}(\mathbf{y})\}$}{
        \For{$s \in \{x : p(x|q)>0\}$}{
            $F_{s,n} \leftarrow - \frac{1}{2\sigma^2}\sum_{i=1}^{n} (y_i - f(s))^2$
        }
    }

    \tcp{Recursion}
    \For{$\ell \in \{2,3,\ldots,m\}$}{
        \For{$n \in \{\ell,\ell+1,\ldots,\mathsf{len}(\mathbf{y})\}$}{
            \For{$s \in \{1,2,\ldots,|\Omega|\}$}{
            $F_{s,n} \leftarrow F_{s,n-1}$\\
                \For{$s' \in 
                \{x : p(s|x) > 0\}$}{
                        $F_{s,n} \leftarrow \mathsf{LSE}(F_{s,n},F'_{s',n-1})$
                    }
                $F_{s,n} \leftarrow F_{s,n} - \frac{1}{2\sigma^2}(y_n - f(s))^2$
            }
        }
        $F' \leftarrow F$
    }

    \tcp{Termination}
    $P \leftarrow -\infty$\\
    \For{$s \in \{1,2,\ldots,|\Omega|\}$}{
        $P \leftarrow \mathsf{LSE}(P, F_{s,\mathsf{len}(\mathbf{y})})$
    }

     \KwRet $P$\\

    \caption{Forward algorithm (Eq. \ref{eq:FA})}
    \label{alg:FA}
  \end{algorithm}

  \begin{algorithm}
    \SetKw{Is}{is}
    \SetKw{Not}{not}
    \SetKw{In}{in}
    \SetKw{Continue}{continue}

    \tcp{Initialisation}
    $F, F' \leftarrow - {\infty}_{1 \times \mathsf{len}(\mathbf{y})}$\\
    \For{$n \in \{1,2,\ldots,\mathsf{len}(\mathbf{y})\}$}{
            $F_{n} \leftarrow - \frac{1}{2\sigma^2}\sum_{i=1}^{n} (y_i - f(s_1))^2$
    }

    \tcp{Recursion}
    \For{$\ell \in \{2,3,\ldots,m\}$}{
        \For{$n \in \{\ell,\ell+1,\ldots,\mathsf{len}(\mathbf{y})\}$}{
                $F_{n} \leftarrow \mathsf{LSE}(F_{n}, F'_{n-1})$\\
                $F_{n} \leftarrow F_{n} - \frac{1}{2\sigma^2}(y_n - f(s_{\ell}))^2$
                
        }
        $F' \leftarrow F$
    }

    \tcp{Termination}
    $P \leftarrow F_{\mathsf{len}(\mathbf{y})}$

     \KwRet $P$\\
    \caption{Conditional forward algorithm (Eq. \ref{eq:CFA})}
    \label{alg:CFA}
  \end{algorithm}

\section{Achievable Information Rates}

So far, we have introduced the noisy nanopore channel for modelling the nanopore sequencer and derived message-passing algorithms to compute APPs for this channel. In this section, we consider how to use these APPs to derive an achievable information rate of the nanopore sequencer from a dataset with $N$ triplets of initial states, inputs and outputs from a nanopore sequencer as  $\mathcal{D} = \{(\mathbf{s}_{i},\mathbf{y}_{i},q_{i})\}_{i=1}^{N}$. Firstly, we give an achievable rate of the nanopore sequencer when {\em mismatched decoding} based on the APPs of the noisy nanopore channel is used. Secondly, we use the DTW algorithm to derive an achievable information rate for any chosen $m$ from a dataset with channel inputs that have arbitrary lengths.

\subsection{Mismatched Decoding}

Consider an initial state $q$, a channel input $\mathbf{s}$, and a channel output $\mathbf{y}$. The {\em information density} (rate) with respect to an arbitrary channel $W$, e.g., the noisy nanopore channel, driven by a uniform source is
\begin{align}
    i_{W}(\mathbf{s};\mathbf{y}|q) = 2 -\frac{1}{m} \log \left(  V(\mathbf{s}|\mathbf{y}, q) \right) 
\end{align}
and an achievable information rate is the mean of information density as $I_W = \mathbb{E}_{U \times W}[i_W(\mathbf{S};\mathbf{Y}|q)]$ where $(\mathbf{S},\mathbf{Y}|q) \sim U \times W$. The nanopore channel $W'$ with APP $V'$ has an achievable rate $I_{W'} = \mathbb{E}_{U \times W'}[i_{W'}(\mathbf{S};\mathbf{Y}|q)]$. However, since the APP $V'$ is unknown we approximate it with $V$ through mismatched decoding. With mismatched decoding of the nanopore channel, the achievable information rate is lower bounded as
\begin{align}
    I^{\mathsf{mismatch}}_{W'} &= 2 -\frac{1}{m}\mathbb{E}_{U \times W'}\left[ \log \left( V(\mathbf{S}|\mathbf{Y}, q) \right)  \right]
\end{align}
where $(\mathbf{S},\mathbf{Y}|q) \sim U \times W'$, since we have the relationship $I_{W'} \geq I^{\mathsf{mismatch}}_{W'}$ where their difference is the divergence rate from $V'$ to $V$. By the law of large numbers, we can estimate $I^{\mathsf{mismatch}}_{W'}$ by averaging the information densities with respect to the noisy nanopore channel $W$ of the dataset $\mathcal{D}$ if $m=\mathsf{len}(\mathbf{s}_i)$ for all $i$.

\begin{remark}
    It is assumed that the initial state $q$ is known, since it is likely that we know the `barcode' that is inserted before the data in a DNA strand to enable parallel sequencing of distinct DNA strands. However, without this side-information, the maximum rate loss would be $\log(4^{\tau})/m = 2\tau/m$ bits/base. For $\tau=5$ and $m=100$, the rate loss is bounded by $0.1$ bits/base.
\end{remark}

\subsection{Variable-length Inputs}

Now, let us derive a lower bound on the achievable rate $I^{\mathsf{mismatch}}_{W'}$, which is for a fixed input length $m$, when the data from the nanopore sequencer has variable lengths that satisfy $\mathsf{len}(\mathbf{s}_i) \geq m$ for all $i$. This relaxes the stronger condition of requiring equal and fixed input lengths as earlier.

Consider an arbitrary data entry $(\mathbf{s}_i,\mathbf{y}_i,q_i)$ from the dataset $\mathcal{D}$. For a desired fixed length input $m$, we use the DTW algorithm on $f(\mathbf{s}_i)$ and $\mathbf{y}_i$ to find the maximum-likelihood segmentation $\tilde{\mathbf{t}}_i = (\tilde{t}_{i,1},\tilde{t}_{i,2},\ldots, \tilde{t}_{i,mL_i})$ and use every $m$-th jump time estimate to imperfectly chop it up into $L_i = \floor{\mathsf{len}(\mathbf{y}_i)/m}$ blocks as $\mathbf{y}_i = \tilde{\mathbf{y}}_{i,1} \circ \tilde{\mathbf{y}}_{i,2} \circ \cdots \circ \tilde{\mathbf{y}}_{i,L_i}$. On the other hand, the true segmentation $\mathbf{t}_i=({t}_{i,1},{t}_{i,2},\ldots, {t}_{i,mL_i})$ can perfectly chop the channel output as $\mathbf{y}_i = \mathbf{y}_{i,1} \circ {\mathbf{y}}_{i,2} \circ \cdots \circ {\mathbf{y}}_{i,L_i}$. This gives an imperfect dataset $\tilde{\mathcal{D}}=\{(\mathbf{s}_{i,j},\tilde{\mathbf{y}}_{i,j},q_{i,j})\}_{i,j}$ and a perfect dataset ${\mathcal{D}}'=\{(\mathbf{s}_{i,j},{\mathbf{y}}_{i,j},q_{i,j})\}_{i,j}$ that satisfies $m=\mathsf{len}(\tilde{\mathbf{y}}_{i,j})=\mathsf{len}({\mathbf{y}}_{i,j})$ for all $i,j$. Furthermore, $\tilde{\mathcal{D}}$ is ``imperfect'' in the sense that each $\tilde{\mathbf{y}}_{i,j}$ was not generated by inputting $\mathbf{s}_{i,j}$ into the nanopore sequencer and reading the output, as is the case in ${\mathcal{D}}'$ with ${\mathbf{y}}_{i,j}$. Observe that, in the presence of side information of every $m$-th jump time, decoding the imperfectly chopped samples with APP
\begin{IEEEeqnarray}{rCl}
V_{\mathsf{dtw}}(\mathbf{s}_i \mid \mathbf{y}_i)
&=&
V(\mathbf{s}_{i,1} \mid \tilde{\mathbf{y}}_{i,1})\,
V(\mathbf{s}_{i,2} \mid \tilde{\mathbf{y}}_{i,2})
\nonumber\\
&& {}\cdots\,
V(\mathbf{s}_{i,L_i} \mid \tilde{\mathbf{y}}_{i,L_i}).
\end{IEEEeqnarray}

is a mismatched decoder for the nanopore channel. Therefore, the mismatched decoding rate using $V_{\mathsf{dtw}}(\mathbf{s}_i|\mathbf{y}_i)$ is an achievable rate that we can estimate as
\begin{align}
    I_{\rm{dtw}} &= 2 - \frac{1}{mN_{\rm{s}}} \sum_{i=1}^{N} \log \left( V_{\mathsf{dtw}}(\mathbf{s}_{i}|{\mathbf{y}}_{i}, q_{i}) \right)\notag\\
    &= 2 - \frac{1}{mN_{\rm{s}}} \sum_{i=1}^{N}\sum_{j=1}^{L_i} \log \left( V(\mathbf{s}_{i,j}|\tilde{\mathbf{y}}_{i,j}, q_{i,j}) \right)
\end{align}
where $N_{\rm{s}}=\sum_{k=1}^{N} L_k$. Note that we assume that the dataset is split into reads from the same nanopore channel, and if there is only one per channel as in the numerical results to come then $N=1$.

\section{ONT Dataset}
Using the techniques developed in the previous section, we study estimates of achievable information rates of the ONT dataset \cite{ONTdataset}. In particular, we use the GM24385 dataset (Chromosome 1) that was created by shotgun sequencing a genome with the PromethION sequencing device that is composed of $2{,}675$ channels and $10{,}700$ R9.4.1 nanopores. This dataset contains raw nanopore signals that correspond to hundreds of reads of base sequences aligned to a reference genome. Since these base sequences are from a genome, their lengths are variable, ranging from hundreds of bases to millions of bases. Within this dataset, we choose $10$ benchmark signals (with read IDs given in Table \ref{tab:ONT_dataset}) out of a total of $117$ that satisfy the following constraints:
\begin{itemize}
    \item Range of number of input bases: 
    $8{,}000 < \mathsf{len}(\mathbf{s}) < 12{,}000.$
    \item Uniform base inputs: 
    $\bigl| N_b(\mathbf{b}) / \mathsf{len}(\mathbf{b}) - 1/4 \bigr| < 0.01,$ for all $b \in \mathcal{B}.$
    \item Maximum mean duration: 
    $\mathsf{len}(\mathbf{y}) / \mathsf{len}(\mathbf{s}) < 20.$
\end{itemize}
The minimum number of input bases is constrained to $8{,}000$ so that there are an abundant number of available signals containing many blocks after chopping. The maximum number of input bases is limited to $12{,}000$, and the maximum mean duration is constrained to $20$, so that the memory requirement of the DTW algorithm is low enough. In addition, this range of lengths of base sequences results in lengths $\sim 10{,}000$, so that we have $\sim 100$ blocks per read in order to be able to average within a read. To support the assumption of uniform input codewords, the base sequence is constrained to be strongly $0.01$-typical with respect to a uniformly distributed base source.

\subsection{Model Parameterisation}
We now parameterise the noisy nanopore channel so that we can decode nanopore signals using the APP $V_{\mathsf{dtw}}$. Since the signals from the genomic dataset are from different nanopore channels, the channel parameterisations would ideally be calibrated to the specific nanopores. Instead, we use models with minimal calibration by using the Scrappie simulator with memory length $\tau=5$. 

The level mapping $f$ for an average nanopore is estimated using levels from Scrappie. These levels could be improved upon by including a calibration step \cite{Vidal2025}---if sufficient training data is available---or by increasing $\tau$ beyond five.  

The mean duration is estimated as
\begin{align}
    \mathbb{E}[K] = \frac{\mathsf{len}(\mathbf{y}_i)}{\mathsf{len}(\mathbf{s}_i)}
\end{align}
which is an accurate estimate since the genomic signals are long. Observe in Table \ref{tab:ONT_dataset} that the mean durations can vary quite a lot, from $10.6$ to $15.7$ samples per base, due to variations in translocation speeds of the motor proteins that function independently of their respective nanopores. Furthermore, this confirms that the channels are indeed not identical. { We note that there is model mismatch in the i.i.d. geometric durations since they are actually non-i.i.d. such that $\mathbb{E}[K]$ depends on the channel state. While considering state-dependent duration parameters could theoretically reduce the segmentation noise, they are challenging to estimate in practice since they require accurate segmentation.} 

{ The measurement noise level is chosen as $\sigma=0.5$, since we found this value approximately gives the highest information rates. The rates were not sensitive to the choice of noise level, however, near the lower bound $\sigma_{\rm dtw}$ the rates were relatively poor.} Observe that this is much higher than the tight lower bound $\sigma_{\mathsf{dtw}}$ in Table \ref{tab:ONT_dataset} due to mismatch relative to the channel model that uses Scrappie levels and {i.i.d. Gaussian noise}. { This is likely due to level mismatches, resulting in a non-i.i.d. Gaussian noise component that depends on the channel state.} The level mismatch could also occur due to the {\em methylation} of some bases in the genomic DNA sequences, temporarily changing the level mapping $f$ for a few percent of the states. Otherwise, the mismatches could be due to variations in the pore construction {that are not considered in the Scrappie levels that are averaged over many nanopores---this could be reduced with a pore-specific calibration step}. 

\subsection{Achievable Information Rates}

Using the discussed model parameterisation, let us now compute $I_{\rm dtw}$ for the ONT dataset to get achievable information rates for nanopore-based DNA storage. It is assumed that each channel-use has an input DNA strand of $m=100$ bases in length, and that the initial state of the nanopore is known. Then, this rate is achievable by coding over many short DNA strands. Note that we do not consider the problem of re-ordering the DNA strands, which has been studied independently \cite{Hossein2016}.

\subsubsection{Mismatched Decoding Degradation}

In Table \ref{tab:ONT_dataset}, the channel ID of each nanopore is given along with its achievable rate $I_{\rm dtw}$, based on a signal from the ONT dataset with a unique read ID. The worst nanopore channel has a rate at least $0.64$ bits/base, and the best nanopore channel has a rate at least $1.18$ bits/base. If we assume that a DNA strand uniformly chooses among these ten channels, the average rate is $0.96$ bits/base. If we use DTW on the blocks to remove $11$ outlier blocks with $\sigma_{\mathsf{dtw}}>0.35$, the average rate goes to $0.99$ bits/base. These blocks likely have sharp changes in the channel conditions that cause a severe channel outage, such as the methylation phenomenon that does not occur in synthetic DNA.

For comparison, the base-sequence for each read ID from the ONT dataset is input into the noisy nanopore channel to generate a synthetic dataset to compute the rate with the APP $V_{\mathsf{dtw}}$ (mismatch due to DTW segmentation and measurement noise) and the rate with the APP $V$ (mismatch due to measurement noise). The measurement noise used to generate this synthetic dataset is set to the lower bound $\sigma_{\mathsf{dtw}}$, such that the resulting rates are slightly optimistic. The average rates are {both} $\approx 1.14$ bits/base, suggesting that the DTW segmentation does not significantly degrade the nanopore channel. In addition, the average degradation compared to the ONT dataset is $0.18$ bits/base. {This level of degradation indicates an upper bound on the performance gain possible with perfect calibration of the channel parameters, since the channel is less mismatched with the synthetic dataset. In addition,} we note that in \cite[Fig. 5(a)]{McBain2024a} the degradation was predicted to be around $0.20$ bits/base due to mismatch in Scrappie levels with $\tau=5$ compared to the Scrappie simulator that assumes $\tau =49$. {This could suggest that the levels are the most critical parameters in the channel model.}

\subsubsection{Channel Quality Scores}

From the information rates, we have observed that the quality of the nanopore channel can vary greatly. When a basecaller is used, this channel quality is typically measured using a Q-score that attempts to predict the block detection error rate of the basecaller. The ONT dataset includes Q-scores from the Guppy basecaller, as given in Table \ref{tab:ONT_dataset}. The Q-scores of the ten signals that we study are representative of a wide range of channel qualities, which are typically between $Q10$ and $Q15$ and only $19\%$ of the signals in the entire dataset are $>Q15$ quality. This suggests that we have included realisations of the worst and the best nanopore channels. Observe that higher Q-scores generally lead to higher rates, confirming that Q-score and rate are equally good measures of channel quality. This is similarly true for DTW distance, however, this requires a known reference signal. Therefore, the Q-score measure or DTW distance may be used to determine the channel quality using a known pilot sequence. This may be used to select particular nanopores capable of high data rates, at the cost of having longer sequencing times. For example, channel $2522$ in Table \ref{tab:ONT_dataset} is not reliable and should be avoided if possible.

 \begin{table*}[]
\centering
\begin{tabular}{@{\extracolsep{4pt}}llccccccccc}
\toprule   
 Ch. ID & Read ID & Num. of blocks & Rate & DTW dist. & Mean duration & Q-score \\ 
\midrule
%
2522&\footnotesize{4da235eb-f848-448a-9f22-3f6238023461}& 116& 0.64& 0.28 & 12.4 & 10.0 \\
682&\footnotesize{58b05ec9-fec7-47ae-b107-564153d0df19}& 111& 0.80& 0.27 & 15.1 & 12.6 \\
1023&\footnotesize{8ea1acc8-92bb-4454-bb71-47af92d56af9}& 105& 0.88& 0.25 & 10.6  & 13.3 \\
2943&\footnotesize{72f77971-83bc-485c-b4b3-8ea2d092a9f2}& 111& 0.95& 0.25 &12.3& 13.9 \\
76&\footnotesize{93e1e23c-d6ab-436c-8461-609226a85474}& 90& 1.00& 0.24 & 11.1& 14.2 \\
764&\footnotesize{7b590d40-18a1-459c-98ed-57be285e6c20}& 94& 1.12& 0.23& 11.6 & 14.6 \\
953&\footnotesize{1fe29082-1543-464c-bd10-ea6dfaad4b74}& 109& 1.09& 0.24 & 11.1 & 14.7 \\
1956&\footnotesize{4b759916-a373-42a7-8001-15f0afd14843}& 100& 1.02& 0.25 & 11.1 & 15.1 \\
522&\footnotesize{7243b2af-6537-4a57-90d0-7280c09e6055}& 113& 1.18& 0.23 & 12.8 & 15.2 \\
2094&\footnotesize{723ffbcb-5626-4f14-a33c-093e32ee7fde}& 113& 0.94& 0.23 & 15.7 & 15.5 \\
\bottomrule
\end{tabular}
\caption{Analysis of achievable information rates in nanopore-based DNA storage using the ONT dataset.} 
\label{tab:ONT_dataset}
\end{table*}

\subsection{Outage rate}\label{sec:outages}
The achievable rate analysis is now extended to the outage rate of the information density. Formally, the outage rate is given by $P_{\rm o}(\gamma) = \mathbb{P}(i_{W}(\mathbf{S}; \mathbf{Y}|q) \leq \gamma)$
for a given threshold $\gamma$ and channel $W$. While not in the scope of this paper, this is closely related to finite block-length coding in DNA storage \cite{Maarouf2022FBL}, since the block error rate at a rate $\gamma$ is essentially determined by how close the outage rate is to $0$ (if it equals $0$, then $\gamma$ is an achievable rate). Furthermore, this is particularly useful for the typical scenario where rate adaptation is not possible, e.g. the DNA encoder (the transmitter) does not know the nanopore it will use but the nanopore sequencer (the receiver) does know it. Therefore, the DNA encoder will use a constant code rate that will result in outages that cause poor block error rates due to statistical variations in the nanopore. By considering the distribution of the information density, this measure provides a deeper characterisation of the nanopore channel than the mean of the information density.

\begin{figure}
    \centering
    \includegraphics[width=1\linewidth]{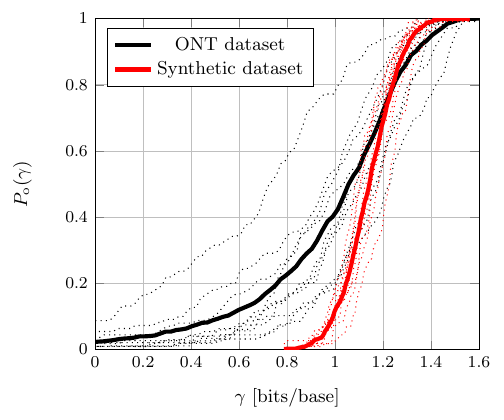}
    \caption{The outage probability of the ONT dataset and the synthetic dataset using the NNC-Scrappie APP decoder.}
    \label{fig:outage_prob}
\end{figure}

Outage rates are given in Fig. \ref{fig:outage_prob} for the ONT dataset (in black) and the synthetic dataset (in red), as used earlier, for the ten nanopore channels (dotted) in Table \ref{tab:ONT_dataset}. In addition, the outage rates are given in thick lines for a randomly chosen nanopore. Firstly, observe that outage rates of the synthetic dataset conform tightly to a Gaussian distribution with mean $1.14$ and variance $0.013$. On the other hand, outage rates of the ONT dataset have heavier left tails skewed towards lower rates. The worst channels have heavier left-tails. For example, a rate of $1$ bit/base approximately achieves an outage probability of $0.1$ for the synthetic dataset but an outage probability of $0.4$ for the ONT dataset, when choosing a random nanopore from the flowcell. Note that the outages for rates below zero are likely due to anomalous outliers that can be removed using a DTW threshold on the blocks as explained earlier.

\section{Conclusion}
 In conclusion, decoding nanopore signals using the NNC-Scrappie decoder in nanopore-based DNA storage can achieve rates of approximately $0.96$ bits per base. The rates of individual nanopore channels vary drastically from $0.64$ bits/base to $1.18$ bits/base, resulting from variations in the nanopore due to defects in the membrane, pore proteins, or electrode connections that are incurred in manufacturing or while in use. However, these lower bounds on achievable rates are likely pessimistic since the DNA storage dataset was imperfectly generated from a genomic dataset. The information-outage analysis highlighted model mismatches when the decoder is applied to the ONT dataset compared to a synthetic dataset, suggesting that a more thorough analysis is needed to determine if the channel model is suitable in the finite block-length regime.
 
 {The achievable rates could be improved with an additional calibration step to update the pore model---the levels in particular---to individual nanopores in the flowcell. In addition, more accurate pore models that consider state-dependent measurement noise and durations could be fruitful for refining the channel model. Deletions could be included in the channel model to make the decoder more robust to backtracking errors. Marginal gains could be achieved by increasing the memory length $\tau$---which would be necessary for the R10.4 Oxford nanopore---however, this may need to be paired with efficient state-space reduction techniques to maintain fast decoding algorithms.}

  The algorithms presented for decoding have simple implementations and could be applied to other datasets when they are available for an updated and possibly more accurate achievable rate and information-outage analysis. { In addition, these algorithms may be applied to protein sequencing by extending the input alphabet and changing the pore model.} 
  
   {Finally, the methods developed in this paper resulted in} new numerical results that provide strong evidence on the applicability of the noisy nanopore channel for modelling nanopore-based DNA storage. {This is an important foundation for future theoretical analyses using pore-based models with noisy duplications.}

\bibliography{refs}
\bibliographystyle{IEEEtran}

\end{document}